\documentclass{article}
\usepackage{spconf,amsmath,graphicx}

\usepackage{cite}
\usepackage{bm}
\usepackage{enumerate}
\usepackage{balance}
\usepackage{float}
\usepackage{array}
\usepackage{multirow}
\usepackage{graphicx}
\usepackage{url}
\usepackage{color}
\usepackage{amssymb}
\usepackage{amsfonts}
\usepackage{amsmath}
\usepackage{extarrows}
\usepackage{graphicx}
\usepackage{amsfonts}
\usepackage{caption}
\usepackage{algorithm}
\usepackage{algpseudocode}
\usepackage{indentfirst}
\usepackage{caption}
\usepackage{esint} 
\usepackage{graphicx}
\usepackage{graphics}
\usepackage{pgfplots}
\usepackage{tikz}
\usepackage{epsfig}
\usepackage{soul}
\usepackage{caption}
\usepackage{gensymb}

\def\ba{{\mathbf a}}

\def\bh{{\mathbf h}}

\def\bn{{\mathbf n}}

\def\by{{\mathbf y}}

\def\bA{{\mathbf A}}

\def\bH{{\mathbf H}}
\def\bI{{\mathbf I}}

\def\bN{{\mathbf N}}

\def\bY{{\mathbf Y}}

\title{Joint Channel and Direction Estimation for Ground-to-UAV Communications Enabled by A Simultaneous Reflecting and Sensing RIS}
\name{Jiguang He$^{\star}$ \qquad Aymen~Fakhreddine$^{\star}$\qquad George C. Alexandropoulos$^{\star \dagger}$}
  
  \address{$^{\star}$Technology Innovation Institute, 9639 Masdar City, Abu Dhabi, United Arab Emirates \\
      $^{\dagger}$Department of Informatics and Telecommunications,\\ National and Kapodistrian University of Athens, 15784 Athens, Greece}

\begin{document}
%
\maketitle
\begin{abstract}
Hybrid Reconfigurable Intelligent Surfaces (HRISs), which are capable of simultaneous programmable reflections and sensing, are expected to play a significant role in future wireless networks, enabling various Integrated Sensing and Communication (ISAC) applications. In this paper, we focus on HRIS-enabled Unmanned Aerial Vehicle (UAV) networks and design the HRIS parameters (phase profile, reception combining, and the power splitting between the two functionalities) for jointly estimating the individual UAV-HRIS and HRIS-base-station channels as well as the Angle of Arrival (AoA) of the Line-of-Sight (LoS) component of the UAV-HRIS channel. We derive the Cram\'er Rao lower bounds for the estimated channels and evaluate the performance of the proposed approach in terms of the channel estimation error and the LoS AoA estimation accuracy, verifying its effectiveness for HRIS-enabled ground-to-UAV wireless communication systems. 
\end{abstract}
\section{Introduction}
\label{sec:intro}
\vspace{-0.2cm}
Reconfigurable Intelligent Surfaces (RISs), commonly in the form of arrays of metamaterials of ultra-low-power response control and without active elements, are lately gaining increased research attention 
for boosting various wireless networking objectives, e.g., spectral and energy efficiencies, as well as localization and sensing~\cite{Huang2018, huang2019holographic, wymeersch2019radio}. However, such passive and computationally-incapable RISs bring considerable difficulty in channel state acquisition. 
In addition, the individual channels in RIS-enabled reflection paths are tightly coupled, which renders the extraction of channel parameters (e.g., paths' directions) intractable. 

Recent research trends are focusing on alternative RIS hardware architectures \cite{Tsinghua_RIS_Tutorial}, including hybrid RISs (HRISs) that are equipped with limited numbers of reception Radio Frequency (RF) chains, enabling simultaneous tunable reflection and sensing (i.e., measurement collection in baseband)~\cite{alexandropoulos2021hybrid,HRIS_Nature}.
These two functionalities are controlled by power splitters which are embedded in each hybrid meta-element or can be deployed per groups of meta-elements. In this regard, the HRIS received signals can be used for channel estimation \cite{Alexandropoulos_oneRFchain,Schroeder2022,lin2021tensor,Zhang2022}, localization \cite{Alexandropoulos2022}, and tracking with affordable cost, power consumption, and computation complexity. 
In addition, HRISs are deemed a promising technology for Integrated Sensing and Communications (ISAC) thanks to their inherent capability to sense/process the impinging signals, improved flexibility, and large-sized aperture~\cite{He2022}. To this end, super-resolution estimation of spatial/angular parameters, e.g., angles of arrival (AoAs), can be realized. When an HRIS operates in a wide band, temporal parameters of the channels (e.g., time of arrival (ToA)) can be estimated with high accuracy. It is noted that sensing tasks are not restricted to the estimation of channel parameters and localization, but also include user absence/presence detection, object recognition, and gesture classification~\cite{Liu2022}.   

In this paper, we consider an HRIS-enabled ground-to-UAV system, where pilot signals are transmitted by the UAV and received by both a Base Station (BS) and the HRIS, enabling the estimation of various channel parameters. In particular, the received signals are jointly processed to estimate the individual UAV-HRIS and HRIS-BS channels as well as the AoAs of the Line-of-Sight (LoS) component of the UAV-HRIS channel. It is noted that this estimation is not feasible with a passive RIS devoid of sensing capabilities. 
We showcase that, by adjusting the HRIS power splitting coefficient, well-balanced performance can be achieved for the estimation of the two individual channels as well as the refinement of the AoA estimation. 
The theoretical Cram\'er Rao Lower Bounds (CRLBs) of the channel estimations are also derived, verifying the effectiveness of the proposed approach and its behavior over different parameter settings.

\textit{Notations}: Bold lowercase and bold capital letters represent matrices and vectors, respectively. $(\cdot)^\mathsf{T}$, $(\cdot)^\mathsf{H}$, and $(\cdot)^{-1}$ denote matrix transposition, Hermitian transposition, and matrix inverse, respectively. $\mathrm{diag}(\ba)$ is a square diagonal matrix with the entries of $\ba$ on its diagonal, $\mathrm{vec}(\bA)$ denotes the vectorization of $\bA$ by stacking its columns on top of one another, $\otimes$ and $\diamond$ are the Kronecker and Khatri-Rao products, respectively, $\mathbb{E}[\cdot]$ is the expectation operator, $\mathbf{0}$ denotes the all-zero vector or matrix, $\bI_{M}$ ($M\geq2$) is the $M\times M$ identity matrix, and $j = \sqrt{-1}$. $\mathrm{Tr}()$ and $\mathrm{Toep}()$ are the trace operator and a Toeplitz matrix formulated by the argument within the brackets, respectively, $\|\cdot\|_2$ denotes the Euclidean norm of a vector and $|\cdot|$ returns the absolute value of a complex number. Finally, notation $\mathcal{CN}(a,b)$ denotes the complex Gaussian distribution with mean $a$ and variance~$b$.

\section{System and Channel Models}
\label{sec:format}
\begin{figure}[t]
	\centering
\includegraphics[width=0.94\linewidth]{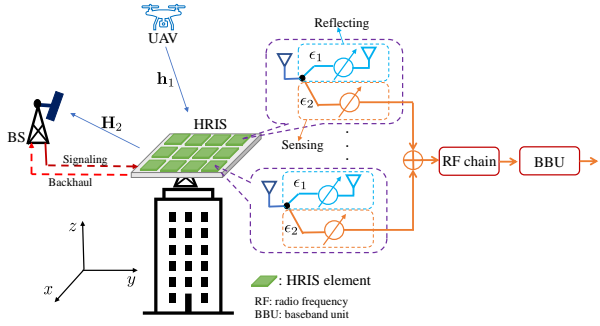}
	\caption{The considered HRIS-enabled ground-to-UAV system including a simultaneous reflecting and sensing metasurface.}
		\label{System_Model}
		\vspace{-0.5cm}
\end{figure}
The considered system model is illustrated in Fig.~\ref{System_Model}, where an HRIS in the form of uniform planar array (UPA) is deployed in parallel to the $x\text{-}y$ plane, enabling communication between a ground BS and a UAV via tunable reflections, while sensing the latter's signal. In particular, by properly designing the HRIS sensing and reflection elements, as well as the power splitting between these two operations, the metasurface can estimate the channel parameters of the HRIS-UAV channel (e.g., the azimuth and elevation AoAs of its line-of-sight (LoS) path), while reflecting the UAV signals to the BS, enabling estimation of the end-to-end link's individual channels (i.e., the UAV-HRIS and HRIS-BS channels). We assume that the HRIS is equipped with a single reception RF chain, the meta-atoms of \cite{HRIS_Nature} for simultaneous reflection and sensing, as well as phase shifters for analog combining \cite{alexandropoulos2021hybrid,Zhang2022}. The received signals at the HRIS are forwarded to the BS via a control link (in- or out-of-band \cite{RISE6G_COMMAG}), and then the BS performs the estimation of the individual channels and AoAs.

\vspace{-0.3cm}
\subsection{Channel Model}
We consider a geometric channel model consisting of angles of departure (AoDs), AoAs, and gains of multiple propagation paths, either LoS or non-line-of-sight (NLoS)~\cite{Cheng2020}. Due to poor scattering, there only exists a limited number of paths connecting any pair of network nodes; this results in spatial sparsity. This nice property is widely used in channel parameters' estimation with low training overhead~\cite{Wang2020,he2020,Wei2021}. The UAV-HRIS channel $\bh_1 \in \mathbb{C}^{M}$ is expressed as follows:
\begin{equation}\label{h_1}
\bh_1 = \sum_{i =1}^{L_1}\frac{e^{-j 2\pi d_{1,i}/\lambda}}{\sqrt{\rho_{1,i}}} \boldsymbol{\alpha}_x(\theta_{1,i},\phi_{1,i}) \otimes \boldsymbol{\alpha}_y(\theta_{1,i},\phi_{1,i}),
\end{equation}
where $L_1$ denotes the propagation paths, $M$ is the number of HRIS elements, $\lambda$ is the wavelength of the carrier frequency, $d_{1,i}$ (in meters) and $\rho_{1,i}$ are respectively the distance and pathloss between the UAV and HRIS, $\theta_{1,i}$ and $\phi_{1,i}$ are the azimuth and elevation AoAs associated with the $i$th path of $\bh_1$, respectively. Without loss of generality, $i=1$ corresponds to the LoS path. 
The array response vectors $\boldsymbol{\alpha}_x(\theta_{1,i},\phi_{1,i})$ and $\boldsymbol{\alpha}_y(\theta_{1,i},\phi_{1,i})$ in \eqref{h_1} can be expressed as~\cite{Tsai2018}:
 \begin{align}
     \boldsymbol{\alpha}_x(\theta_{1,i},\phi_{1,i}) =& \Big[1, e^{j \frac{2\pi d_{1,x}}{\lambda} \cos(\theta_1) \sin(\phi_{1,i})},\nonumber \\
    & \ldots, e^{j \frac{2\pi d_{1,x}}{\lambda} (M_x -1) \cos(\theta_{1,i})\sin(\phi_{1,i})} \Big]^{\mathsf{T}}, \nonumber\\
       \boldsymbol{\alpha}_y(\theta_{1,i}, \phi_{1,i}) =& \Big[1, e^{j \frac{2\pi d_{1,y}}{\lambda} \sin(\theta_{1,i}) \sin(\phi_{1,i}) },\nonumber \\
    & \ldots, e^{j \frac{2\pi d_{1,y}}{\lambda} (M_y -1) \sin(\theta_{1,i}) \sin(\phi_{1,i})} \Big]^{\mathsf{T}},\nonumber
    \vspace{-0.5cm}
 \end{align}
where $M_x$ and $M_y$ are the number of HRIS elements across the $x$-axis and $y$-axis, respectively, satisfying $M_x M_y = M$, and $d_{1,x}$ and $d_{1,y}$ denote their corresponding inter-element spacings. Similar to \eqref{h_1}, the HRIS-BS channel $\bH_2 \in \mathbb{C}^{N \times M}$, including $L_2$ propagation paths, is expressed as follows:
\begin{equation}
    \bH_2 = \sum_{i =1}^{L_2} \frac{e^{-j 2\pi d_{2,i}/\lambda}}{\sqrt{\rho_{2,i}}} \boldsymbol{\alpha}_z(\varphi_{2,i})  \Big[\boldsymbol{\alpha}_x(\theta_{2,i},\phi_{2,i}) \otimes \boldsymbol{\alpha}_y(\theta_{2,i},\phi_{2,i})\Big]^\mathsf{H}, \nonumber
\end{equation}
where the array response vector $\boldsymbol{\alpha}_z(\varphi_{2,i})$ is given by:
\begin{equation}
     \boldsymbol{\alpha}_z(\varphi_{2,i}) = \Big[1, e^{j \frac{2\pi d_{2,z}}{\lambda}  \cos(\varphi_{2,i}) }, \ldots, e^{j \frac{2\pi d_{2,z}}{\lambda} (N -1) \cos(\varphi_{2,i})} \Big]^{\mathsf{T}}, \nonumber
\end{equation}
with $N$ being the number of BS antennas. $\varphi_{2,i}$, $\theta_{2,i}$, and $\phi_{2,i}$ are the elevation AoA, as well as the azimuth and elevation AoDs associated with $\bH_2$'s $i$th path, respectively. The other parameters are defined similarly to those appearing in~\eqref{h_1}.  

\vspace{-0.3cm}
\subsection{Signal Model}
Assuming that the UAV transmits the pilots $s_k$ with $\mathbb{E}\{|s_k|^2\} = P$ during each $k$th time instant, the received signal at the HRIS can be expressed as follows:
\begin{equation}\label{y_R_k}
    y_{\text{R},k} = \epsilon_2 \boldsymbol{\omega}_{2,k}^\mathsf{H}\bh_1 s_k + n_{\text{R},k},
\end{equation}
where $\boldsymbol{\omega}_{2,k}\in \mathbb{C}^{M}$ includes the configuration of the sensing phase shifters, 
$n_{\text{R},k}$ is the additive white Gaussian noise (AWGN) distributed as $\mathcal{CN}(0, \sigma_\text{R}^2)$, and $\epsilon_2$ controls the power splitting for sensing. In this paper, we consider the same power splitting coefficient for all HRIS elements. Note that element-wise power splitting coefficient is also feasible with increased complexity. In addition, the received signal during the $k$th time instant at the BS is given by:
\begin{equation}\label{by_B_k}
    \by_{\text{B},k} = \epsilon_1 \bH_2 \boldsymbol{\Omega}_{1,k} \bh_1 s_k + \bn_{\text{B},k},
\end{equation}
where $\boldsymbol{\Omega}_{1,k}\triangleq\mathrm{diag}(\boldsymbol{\omega}_{1,k})$ includes the HRIS reflection coefficients $\boldsymbol{\omega}_{1,k}\in \mathbb{C}^{M}$, the AWGN $\bn_{\text{B},k}$ is distributed as $\mathcal{CN}(0, \sigma_\text{B}^2\mathbf{I}_N)$, and $\epsilon_1$ controls the power splitting for reflection. We have the following conditions for the power splitting coefficients $\epsilon_1$ and $\epsilon_2$: \textit{i}) $\epsilon_1, \epsilon_2 \in[0,1]$; and \textit{ii}) $\epsilon_1^2+\epsilon_2^2 =1$. For the reflection and sensing coefficients holds $|[\boldsymbol{\omega}_{1,k}]_{m}|=|[\boldsymbol{\omega}_{2,k}]_{m}|=1$ $\forall$$m=1,2,\ldots,M$ due to the HRIS hardware constraints \cite{Zhang2022}.

By collecting the received signals across $K$ time instants, we can derive the expressions for the measurements: 
\begin{align}\label{by_R}
    \by_\text{R} & = \sqrt{P}\epsilon_2 \boldsymbol{\Omega}_{2}^\mathsf{H}\bh_1  + \bn_{\text{R}}, \\
    \bY_\text{B} &=\sqrt{P} \epsilon_1 \bH_2  \mathrm{diag}(\bh_1)\boldsymbol{\Omega}_{1} + \bN_\text{B},\label{bY_B}
\end{align}
where $\boldsymbol{\Omega}_{1} \triangleq [\boldsymbol{\omega}_{1,1},\ldots,\boldsymbol{\omega}_{1,K}]$ and $\boldsymbol{\Omega}_{2} \triangleq [\boldsymbol{\omega}_{2,1},\ldots,\boldsymbol{\omega}_{2,K}]$, $\by_\text{R} \triangleq[y_{\text{R},1},\ldots, y_{\text{R},K}]^\mathsf{T}$, $\bn_\text{R} \triangleq[n_{\text{R},1},\ldots, n_{\text{R},K}]^\mathsf{T}$, $\bY_\text{B} = [\by_{\text{B},1},\ldots,\by_{\text{B},K}]$, $\bN_\text{B} \triangleq [\bn_{\text{B},1},\ldots,\bn_{\text{B},K}]$, and $s_k$ was set to $\sqrt{P}$ without loss of generality. We then perform the vectorization $\by_\text{B} \triangleq \mathrm{vec}(\bY_\text{B})$, which using the definition $\bn_\text{B} \triangleq \mathrm{vec}(\bN_\text{B})$, yields:
\begin{equation}\label{by_B}
    \by_\text{B} = \sqrt{P} \epsilon_1 (\boldsymbol{\Omega}_{1}^\mathsf{T}  \diamond \bH_2) \bh_1 + \bn_\text{B}.
\end{equation}
Finally, the concatenated vector $\by_\text{RB} \triangleq [\by_\text{R}^\mathsf{T}, \by_\text{B}^\mathsf{T} ]^\mathsf{T}$ is deduced: 
\begin{equation}\label{by_RB}
        \by_\text{RB} = \sqrt{P} \bA_\text{RB} \bh_1+ \bn_\text{RB}, 
\end{equation}
where definition $\bA_\text{RB} \triangleq [\epsilon_2 \boldsymbol{\Omega}_{2};\;\epsilon_1 (\boldsymbol{\Omega}_{1}^\mathsf{T}  \diamond \bH_2)^\mathsf{H} ]^\mathsf{H}$ and $ \bn_\text{RB} \triangleq [\bn_\text{R}^\mathsf{T},  \bn_\text{B}^\mathsf{T}]^\mathsf{T}$ were used.

In this paper, we are interested in the estimation of $\bh_1$ and $\bH_2$ using the measurements in~\eqref{by_R} and~\eqref{bY_B}. In particular, we focus on iterative refining the estimates for these two individual channels as well as the direction (i.e., LoS AoAs) of the UAV with respect to the HRIS.


\section{Proposed Channel and AoA Estimation}
Leveraging the HRIS mode of operation and the availability of the received signals at the HRIS via the BS-HRIS control link, we first perform channel estimation using regularized optimization and atomic norm minimization~\cite{Yang2016IT, he2020anm}. In particular, we commence by estimating $\bh_1$ as follows: 
\begin{align}\label{ANM1}
\hat{\bh}_1 \triangleq \arg \min_{\bh_1 \in \mathbb{C}^{M},\; \mathcal{B}} &\mu_1 \|\bh_1\|_{\mathcal{A}_1} + \frac{1}{2}\|\by_\text{R} - \sqrt{P}\epsilon_2 \boldsymbol{\Omega}_{2}^\mathsf{H}\bh_1\|_2^2, \nonumber\\
&\text{s.t.} \;\begin{bmatrix} \mathrm{Toep}(\mathcal{U}_1)  & \bh_1\\
\bh_1^{\mathsf{H}}& t
\end{bmatrix} \succeq \mathbf{0},
\end{align}
where $\mu_1 \propto \sigma_\text{R} \sqrt{ M \log(M)}$ is the regularization term of the atomic norm penalty, $\mathcal{B}\triangleq\{\mathcal{U}_1 \in \mathbb{C}^{M_{x} \times M_{y} }, t\in \mathbb{R}\}$, $\mathcal{A}_1 \triangleq \{\boldsymbol{\alpha}_x(x_1, x_2) \otimes \boldsymbol{\alpha}_y(x_1,x_2), x_1 \in [0, \pi], x_2 \in [-\pi/2, \pi/2]  \}$ is the atomic set, and $\|\bh_1\|_{\mathcal{A}_1} = \mathrm{inf}_{\mathcal{B}}\Big\{\frac{1}{2M_{x} M_{y}} \mathrm{Tr}(\mathrm{Toep}(\mathcal{U}_1)) + \frac{t}{2}\Big\}$ is the atomic norm. Then, we similarly estimate $\bH_2$ as:
\begin{align}\label{ANM2}
\hat{\bH}_2 \triangleq& \arg \min_{\bH_2 \in \mathbb{C}^{N \times M},\; \mathcal{G}} \mu_2 \|\bH_2\|_{\mathcal{A}_2} \nonumber\\
&+ \frac{1}{2}\|\bY_\text{B} -\sqrt{P} \epsilon_1 \bH_2  \mathrm{diag}(\hat{\bh}_1)\boldsymbol{\Omega}_{1}\|_2^2, \nonumber\\
&\hspace{0.5cm}\text{s.t.} \;\begin{bmatrix} \mathrm{Toep}(\mathcal{U}_{2,1})  & \bH_2\\
\bH_2^{\mathsf{H}}& \mathrm{Toep}(\mathcal{U}_{2,2})
\end{bmatrix} \succeq \mathbf{0},
\end{align}
where $\mu_2 \propto \sigma_\text{B} \sqrt{ MN \log(MN)}$ is the regularization term, $\mathcal{G}\triangleq\{\mathcal{U}_{2,1} \in \mathbb{C}^{N \times N },\, \mathcal{U}_{2,2} \in \mathbb{C}^{M\times M}\}$, $\mathrm{Toep}(\mathcal{U}_{2,1})$ and $\mathrm{Toep}(\mathcal{U}_{2,2})$ are 2-level Toeplitz matrices~\cite{Tsai2018}, $\mathcal{A}_2 \triangleq  \Big\{\boldsymbol{\alpha}_z(x_4) \\ \big[\boldsymbol{\alpha}_x(x_5,x_6) \otimes \boldsymbol{\alpha}_y(x_5,x_6)\big]^\mathsf{H}, x_5 \in [0, \pi], x_4,x_6 \in [-\pi/2, \pi/2]  \Big\}$ is the atomic set, and $\|\bH_2\|_{\mathcal{A}_2} = \mathrm{inf}_{\mathcal{B}}\Big\{\frac{1}{2N} \mathrm{Tr}\big(\mathrm{Toep}(\mathcal{U}_{2,1})\big) + \frac{1}{2M} \mathrm{Tr}\big(\mathrm{Toep}(\mathcal{U}_{2,2})\big)\Big\}$ is the atomic norm. The latter optimization problems can be solved using the approach in \cite{he2020anm}.

The $\bH_2$ estimation obtained from \eqref{ANM2} can be substituted into~\eqref{by_RB} to refine the estimation for $\bh_1$ via a similar optimization to \eqref{ANM1}. In fact, this procedure can be performed iteratively to improve channel estimation, however, at the cost of additional computational complexity. The overall proposed estimation approach is summarized in Algorithm~\ref{alg:proposed_JSP}. As shown, the final $\hat{\bH}_2$ and $\hat{\hat{\bh}}_1$ are derived in steps $2$ and $3$, respectively, whereas step $4$ extracts the AoAs of the LoS component of $\bh_1$. For this estimation, we have used root MUltiple SIgnal Classification (MUSIC)~\cite{Barabell1983}. Note that the estimates for the individual channels can be used to design the HRIS reflection coefficients in closed form \cite{Moustakas_RIS}; hence, optimizing the HRIS for UAV-BS communications. In addition, the direction estimation can deployed for localizing the UAV \cite{Alexandropoulos2022}. 
\begin{algorithm}[!t]
\caption{Joint Channel and AoA Estimation}\label{alg:proposed_JSP}
\begin{algorithmic}[1]
\State \textbf{Initial $\bh_1$ estimation}: Obtain $\hat{\bh}_1$, using~\eqref{by_R}, via \eqref{ANM1}.
\State \textbf{$\bH_2$ estimation}: Use $\hat{\bh}_1$ in~\eqref{bY_B} and obtain $\hat{\bH}_2$ via \eqref{ANM2}.
\State \textbf{Final $\bh_1$ estimation}: Substitute $\hat{\bH}_2$ into~\eqref{by_RB} and obtain $\hat{\hat{\bh}}_1$, similar to $\hat{\bh}_1$, via \eqref{ANM1}.
\State \textbf{Parameter estimation for $\bh_1$}: Obtain the estimation of the AoA of the LoS component of $\bh_1$ using $\hat{\hat{\bh}}_1$ and~\cite{Barabell1983}.
\end{algorithmic}
\end{algorithm}
\vspace{-0.3cm}
\subsection{CRLB Analysis}
The theoretical Mean Square Error (MSE) for the estimates of $\bh_1$ and $\bH_2$ are given as follows~\cite{Tichavsky1998}:
\begin{align}\label{CRLB1}
    \mathbb{E} \big\{ \|\bh_1-\hat{\hat{\bh}}_1\|_2^2| \bH_2\big\} & \geq \mathrm{Tr}\bigg\{\bigg[\Re\Big(\frac{2P}{\sigma^2}\bA_\text{RB} \bA_\text{RB}^\mathsf{H} \Big)\bigg]^{-1}\bigg\}, \\
    \mathbb{E} \big\{ \|\bh_2-\hat{\bh}_2\|_2^2| \bh_1\big\} & \geq \mathrm{Tr}\bigg\{\bigg[\Re\Big(\frac{2P}{\sigma^2}\epsilon_1^2\bar{\boldsymbol{\Omega}}_{1}\bar{\boldsymbol{\Omega}}_{1}^\mathsf{H}\Big)\bigg]^{-1}\bigg\},\label{CRLB2}
\end{align}
where $\sigma^2 = \sigma^2_{\text{R}}=\sigma^2_{\text{B}}$,  $\bar{\boldsymbol{\Omega}}_{1} \triangleq (\boldsymbol{\Omega}_{1}^\mathsf{T}\mathrm{diag}(\bh_1) \otimes \bI)$, $\bh_2 \triangleq \mathrm{vec}(\bH_2)$, and $\hat{\bh}_2 \triangleq \mathrm{vec}(\hat{\bH}_2)$. We have assumed perfect knowledge of $\bH_2$ and $\bh_1$ in~\eqref{CRLB1} and~\eqref{CRLB2}, respectively, relaxing the tightness of the bounds. Nevertheless, they can still be used for studying the performance of the proposed channel estimation approach. Note that both MSE performances depend on the power splitting between the HRIS operations.

\section{Numerical Results}
In this section, we assess the role of the HRIS power splitting coefficient on the joint channel and AoA estimation performance. We particularly evaluate the normalized MSE (NMSE) of both channel estimates as well as the LoS AoA estimation in the UAV-HRIS channel. 
In the results that follow, we have considered the following parameters' setting: $M = 6 \times 6$, $N = 4$, $T = 500$, $L_1=L_2=2$, and $K = 32$. The locations of the UAV, HRIS, and BS were at the points $(6,9,10)$, $ (3,4,2)$, and $(2.5,3.5,1.5)$, respectively, from which the angular parameters and distances associated with the LoS path can be calculated. The angular parameters in the NLoS paths were set to $\pi/4$ away from those of the respective LoS ones. The pathloss $\rho_{i,1}$, with $i=1$ and $2$, was modeled as $\rho_{i,1} =10^{3.245} d_{i,1}^2 f_c^2$~\cite{Rappaport2013} with $f_c$ (in GHz) being the carrier frequency, which was set to $3.5$ GHz. 
The average powers of the LoS and NLoS paths at each hop were varied for $10$ dB. Finally, during the pilot transmission phase, $\boldsymbol{\Omega}_1$ and $\boldsymbol{\Omega}_2$ were selected from the first $K$ columns of the $M \times M$ discrete Fourier transform (DFT) matrix.    

\begin{figure}[t]
	\centering
\includegraphics[width=0.99\linewidth]{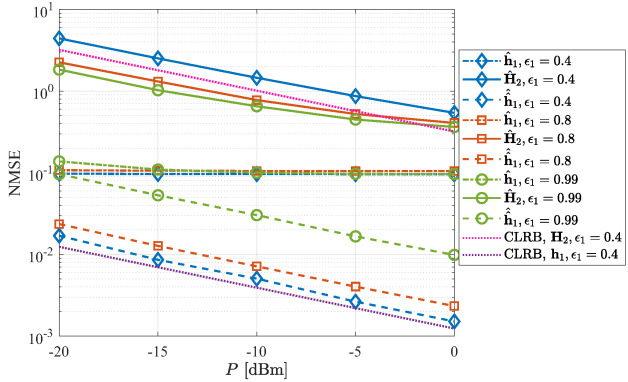}
	\caption{The effect of the HRIS power splitting coefficient $\epsilon_1$ on the NMSE performance of both channel estimates.}
		\label{NMSE_results}
		\vspace{-0.5cm}
\end{figure}

The NMSE performance of the proposed joint channel and AoA estimation is illustrated in Fig.~\ref{NMSE_results} versus the transmit power $P$ for different $\epsilon_1$ values. The CRLB results for the case of $\epsilon_1 = 0.4$ are also included. As shown, when $\epsilon_1$ increases, the estimation of $\bH_2$ improves, while that of $\bh_1$ degrades. This verifies that, by well controlling the HRIS power splitting coefficient, we can achieve the optimal performance tradeoff between the two individual channel estimates. 

\begin{figure}[t]
	\centering
\includegraphics[width=0.99\linewidth]{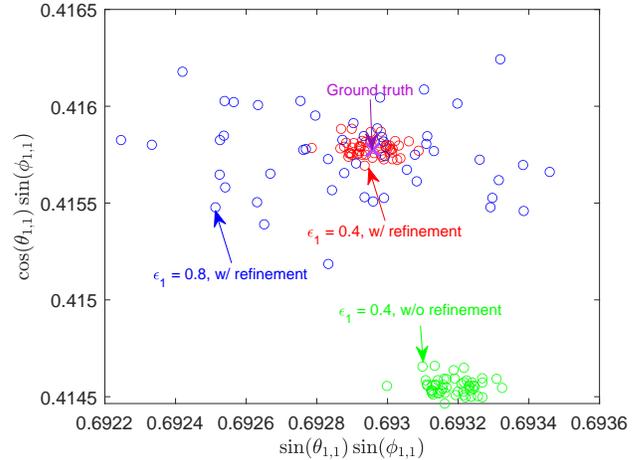}
	\caption{Performance enhancement of the LoS AoA estimation for the UAV-HRIS channel by joint processing of the received signals, considering $P = -10$ dBm and $\epsilon_1 = 0.4$ and $0.8$. }
		\label{DoA_results}
		\vspace{-0.5cm}
\end{figure}

In Fig.~\ref{DoA_results}, we demonstrate the performance enhancement of the LoS AoA estimation for the UAV-HRIS channel via the proposed joint channel and AoA estimation approach (marked as ``w/ refinement" in the figure) by considering $50$ trials/realizations (each circle refers to the AoA estimation and is colored according to the case used). The scheme of considering only the estimates obtained from steps $1$ and $2$ in Algorithm~\ref{alg:proposed_JSP} is also included as a benchmark (marked as ``w/o refinement"). As depicted, higher-precision LoS AoA estimation can be achieved with the proposed joint estimation approach compared to the considered benchmark. By inspecting Fig.~\ref{DoA_results} for $\epsilon_1 = 0.8$ and $0.4$, we conclude that the former value yields a poorer LoS AoA estimation, which is consistent with the channel estimation results from Fig.~\ref{NMSE_results}.

\section{Conclusion}
In this paper, we have studied joint channel and AoA estimation in a HRIS-enabled ground-to-UAV communication system. We have particularly investigated the effect of the HRIS power splitting coefficient on the estimations for the individual channels and the AoAs of the LoS path of the UAV-HRIS link. By jointly processing the received signals from the HRIS and BS and controlling power splitting, we were able to improve the tradeoff between the two channels' estimates and the refinement of the UAV direction finding process.

\section{Acknowledgement}
\vspace{-2mm}
Prof. G.~C.~Alexandropoulos acknowledges the support by the EU H2020 RISE-6G project under the grant number 101017011.

 \balance
\vfill\pagebreak

\bibliographystyle{IEEEbib}
\bibliography{IEEEabrv,refs}

\end{document}